# Experimental radar absorption in high-filling factor magnetic composites

Jaume Calvo-de la Rosa[1,2,*], Marc Vazquez-Aige[1], Paula Pérez[3], Laura Medina[3], Pilar Marín[4,5], Jose Maria Lopez-Villegas[2,6], Javier Tejada[1]

[1] *Departament de Física de la Matèria Condensada, Universitat de Barcelona, Martí i Franquès 1, 08028 Barcelona, Spain*

[2] *Institut de Nanociència i Nanotecnologia (IN2UB), Universitat de Barcelona, 08028 Barcelona, Spain*

[3] *PREMO S.L., 29590 Malaga, Spain*

[4] *Instituto de Magnetismo Aplicado (IMA-UCM-ADIF), 28230 Madrid, Spain*

[5] *Departamento de Física de Materiales, Facultad de Físicas, Universidad Complutense de Madrid (UCM), 28040 Madrid, Spain*

[6] *Departament d'Enginyeria Electrònica i Biomèdica, Universitat de Barcelona, 08028 Barcelona, Spain*

*\* Corresponding author: jaumecalvo@ub.edu*

## ABSTRACT

The electromagnetic properties and microwave absorption capabilities are studied in an anechoic chamber under real radar conditions for mono and bilayer composite samples consisting of a polymeric matrix and a magnetic powder filler, either metallic or ceramic. The effect of the filler type and the filling factor is investigated. The results demonstrate exceptional broadband microwave absorption, making these materials highly suitable for stealth technology applications. The experimentally measured absorptions reach -40 dB, while model-based predictions suggest that these systems could overpass the barrier of -50 dB. The experimental results are supported by models, both for single and bilayer systems.

## 1. Introduction

Electromagnetic waves, and more specifically microwaves, are present in our everyday live. Telecommunication, wireless connection, remote sensing, or positioning systems for the automotive industry are just a few examples [1], [2], [3], [4], [5], [6]. This continuous flow of microwaves has had a main role in the improvement of nowadays connectivity and the subsequent technological and social advancement. However, while this connectivity may have an exceptional positive impact on society, it also creates new challenges in terms of security [7], [8], [9]. The protection of sensitive systems from remote undesired control (such a hacking) or attacks becomes vital in a world that tends to be every day more automatized and electronically controlled.

Stealth technology aims to prevent electromagnetic radiation to reflect from objects, spaces or systems in general [10], or even to avoid the radiation to penetrate these bodies. Different are the approaches that may be used to avoid the radiation from passing though, but radar-absorbing materials (RAMs) are a key agent in this mission. These are materials which are designed in such a way that they are capable to absorb most of the microwave radiation. The threshold to consider a material as a RAM is typically set to -10 dB, which means that 90% of the power received is attenuated by the material [10], [11], [12]. This means that only a small fraction of the radiation is transmitted or reflected. Today one might easily find -20 dB or -30 dB RAMs in literature [13], [14], which means that 99% and 99.9%, respectively, of the radiation is absorbed. Materials with these properties have a direct impact both for civil and military purposes. However, these materials tend to absorb at specific frequencies, but it is less common to find materials capable of absorbing along a wide range of frequencies. Reaching broadband absorption is one of the main necessities that need to be solved to avoid radiation reflecting independently of the incoming wavelength.

There exist RAMs of different natures and with different compositions. Despite one might also find cases where the absorption mechanism is motivated by the dielectric behavior of the materials [15], [16], [17], [18], one of the most promising strategies is to fabricate composite systems made of a dielectric matrix filled with magnetic particles. This double capacity helps of interacting and attenuating the two components of an electromagnetic wave. The dielectric matrix uses to be polymeric, which (in addition to its dielectric behavior) are cost-effective and lightweight materials. The magnetic fillers may be either ceramic or metallic. Metallic soft magnetic materials are interesting candidates due to their strong magnetic moment, and high magnetic susceptibility and permeability at high frequencies [19], [20]. On the other hand, ceramic materials have an interesting combination of properties. Despite their permeability is lower than metals, they are chemically more stable, cheaper and easier to produce, and have a lower density compared to metals. Focusing on the case of ferrites, which is the most common type of ceramic magnet, there is room to modify its chemical composition and crystalline structure, leading to a large variety of functional properties. These modifications have even allowed finding random magnetic properties in barium hexaferrite-type materials recently [21].

Most of the needs that stealth technology has, especially for radar applications, cannot be solved with high absorption materials in the bulk form. Contrary, it is needed to develop materials that can be processed in the form of thin layers (a few millimeters) to cover surfaces and spaces [8]. Thus, there is also a scientific and technological challenge in synthesizing and processing materials in that shape with good dimensional control and regularity. When composites are prepared, the filler material needs to be homogeneously dispersed as well.

In this paper, we demonstrate the radar absorption capacities of a set of ~1.7 mm thick layer-type composite materials. The electromagnetic properties are initially measured, and the radar absorption capacities are both experimentally measured and compared to models. The recent model [22] is used to model bilayer systems and explore the throughput of the combination of more than one layer, which has been proven as a promising strategy.

## 2. Materials and methods

### a. <u>Materials and sample preparation</u>

All the materials used in this work have been fabricated and characterized by the company PREMO S.L. In the context of an increasingly sustainability-focused strategy, we used soft ferrites waste from the manufacturing processes of magnetic components, as filler. For this purpose, this scrap first underwent a recycling process. For the optimization of the particle size distribution of recycled ferrite powder, an LMC100 jaw crusher was first used to reduce the scrap size to 1-5 mm. Before the next step, the material was submitted for removal of plastic traces treatment. We made an optimization of the subsequent milling process conditions. The powder was micronized by using a Retsch PM100 planetary ball mill under for 8 hours at 300 rpm, by using 30 mm diameter balls, and a fraction of ball powder ratio of 30/1. Given that it resulted in a relatively wide particle size distribution, we used an AS 200 basic vibratory sieve shaker of Retsch to separate the powder into different particle size fractions, using sieves ranging from 800 microns to less than 40 microns.

On the other hand, we also used $Fe_{6.6}Si$ powder, with a mean particle size of 33.5μm. This is a commercial material provided by the company PREMO S.L.

For the preparation of our composite samples, we used commercial standard polydimethylsiloxane (PDMS) as the polymeric base and two different magnetic fillers, $Fe_{6.6}Si$ with a particle size fraction between 33.5 – 50 μm and powder of recycled Mn-Zn ferrite in two different particle size fractions, 40 – 64 μm and < 40 μm.

During its preparation process, there are two different components (A and B) in liquid state, which contain a percentage of magnetic filler mixed with PDMS. In our case, component A is the resin, and component B is the hardener. Components A and B are separately prepared and then mixed in a 1:1 ratio by weight. The obtained composite was deposited on top of a 25cm × 25cm polyester sheet of 50 μm thickness.

Table I below lists the characteristics of the 9 functional samples that we prepared for this study. In addition to these, we also fabricated un-loaded samples consisting only of the polyester sheet and another one made of the polyester sheet plus a pure PDMS deposition on top, which worked as references. For the functional sheets we used the three abovementioned magnetic fillers and, with each of them, we prepared samples with three different weight filling factors ($ff_W$), leading to a total of 9 sheets.

**Table I.** List and characteristics of all the prepared sheets.

| # | Sample ID | Matrix | Filler | *Particle size (μm)* | *$ff_W$ (%)* |
|---|---|---|---|---|---|
| 1 | FeSi_85 | PDMS | $Fe_{6.6}Si$ | 33.5 - 50 | 85.42 |
| 2 | FeSi_64 | PDMS | $Fe_{6.6}Si$ | 33.5 - 50 | 64.08 |
| 3 | FeSi_43 | PDMS | $Fe_{6.6}Si$ | 33.5 - 50 | 42.73 |
| 4 | MnZnF_L_85 | PDMS | Mn-Zn ferrite | 40 - 64 | 85.42 |
| 5 | MnZnF_L_64 | PDMS | Mn-Zn ferrite | 40 - 64 | 64.08 |
| 6 | MnZnF_L_43 | PDMS | Mn-Zn ferrite | 40 - 64 | 42.73 |
| 7 | MnZnF_S_85 | PDMS | Mn-Zn ferrite | < 40 | 85.42 |
| 8 | MnZnF_S_64 | PDMS | Mn-Zn ferrite | < 40 | 64.08 |
| 9 | MnZnF_S_43 | PDMS | Mn-Zn ferrite | < 40 | 42.73 |

The terminology used to define the samples follows the structure: *filler_$ff_W$*. For the case of Mn-Zn ferrite, we also specify, after the filler name, the size of the distribution ("*S*" refers to small, while "*L*" refers to large).

b. Electromagnetic and microwave measurements

The electromagnetic characterization of the sheets at the microwave frequency range was performed by using a Keysight E5071C ENA Series Network Analyzer coupled to a couple of X-band horn antennas, in a two-port measurement configuration. The two antennas were spaced 40 cm, being the sheet holder placed in the center in perfect alignment. An S2P measurement was done, recording the corresponding four complex *S*-parameters ($S_{11}$, $S_{21}$, $S_{12}$ and $S_{22}$) in each measurement. The data was recorded along 1601 points between 2 and 18 GHz, to ensure that all the X-band frequency range was covered. After the proper calibration and analysis of the antenna performance, only data above 8 GHz was finally used.

The system and the cables were initially calibrated through a standard E-Cal method, and additional short/thru reference measurements were done for further calibration of the detected signals. The measured signals were initially filtered (to ensure that only signals coming back from the sample interaction were considered) and the reference short and thru signals were used for doing the proper port-extensions of the signal and accounting for the dissipation effect of the antennas, respectively. Nicolson-Ross-Weir (NRW) [23] method was used to extract the complex magnetic permeability ($\widehat{\mu_r} = \mu_r' - j\mu_r''$) and permittivity ($\widehat{\varepsilon_r} = \varepsilon_r' - j\varepsilon_r''$) from the *S*-parameters.

The experimental far-field reflection loss ($R_L$) was conducted in an anechoic chamber using two EMCO 3160-07 horn antennas (measuring 1601 points in the frequency range between 0.5 and 18 GHz), positioned under far-field conditions relative to the optimized absorbing paint. Under these conditions, the transmission of electromagnetic waves can be considered as plane waves. The sample was fixed on a metallic plate, to ensure total reflection. Additionally, the anechoic chamber is equipped with electromagnetic radiation absorbers and plates to prevent edge effects. The electromagnetic signal is controlled and analyzed by an Agilent E8362B PNA Microwave Network Analyzer, which is calibrated under free-space conditions. The scattering coefficient $S_{21}$ is measured directly, the real and imaginary parts of the reflection coefficient ($R_0$) are obtained, and the reflection loss coefficient is represented as $R_L = 20\log(|R_0|)$, in dB scale. A schematic representation of the two systems is presented in Figure 1.

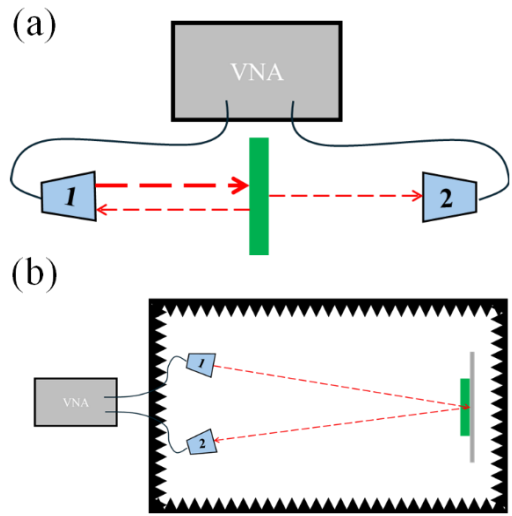

**Figure 1.** Schematic representation of the two microwave measuring systems: (a) antennas in transmission mode for the measurement of the electromagnetic parameters and (b) antennas in reflection mode inside the anechoic chamber for the measurement of the reflection loss.

3. **Results and discussion**

Both samples were initially chemically and structurally characterized to validate their crystal structure. Laser-Induced Breakdown Spectroscopy (LIBS) and x-ray diffraction (XRD) confirmed the pure composition of both samples. Figure 2 provides the XRD pattern obtained for both samples, which reveals a pure $Fe_3Si$ composition for the FeSi sample. In the second case, small traces of pure Fe are identified, though the dominant phase corresponds to the spinel Mn-Zn ferrite. XRD also revealed that the recycled Mn-Zn ferrite powder, after the corresponding milling processes, has an average crystallite size of 33.5μm. Scanning Electron Microscopy (SEM) allowed us to observe that both samples consist of roughly spherical particles of a few tens of micrometers in diameter.

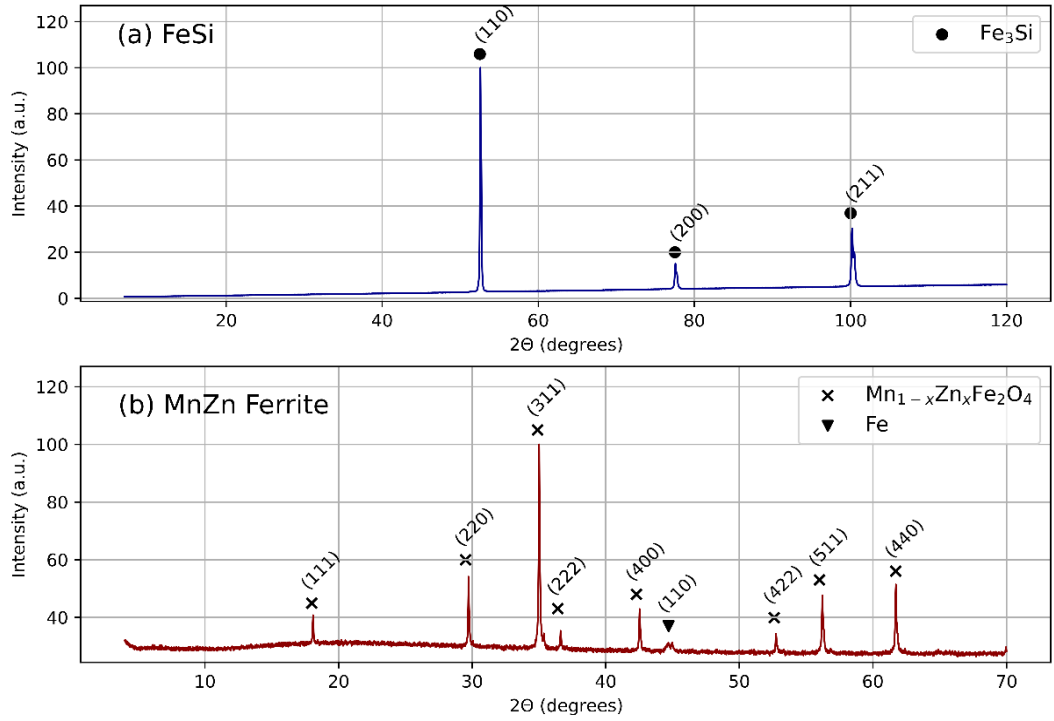

**Figure 2.** XRD pattern obtained for the (a) FeSi and (b) MnZn ferrite samples. Each diffraction peak is identified with the crystal phase (symbol) and the Miller indices (*hkl*) of the corresponding family of planes.

To start with the functional analysis, let us focus on single-layer systems, i.e. the measurement of each sample listed in Table I. The transmitted and reflected signals obtained by the two-ports measurements using the horn antennas were processed through the NRW method to deduce the electromagnetic response of each sample. The obtained responses are shown in Figure 3.

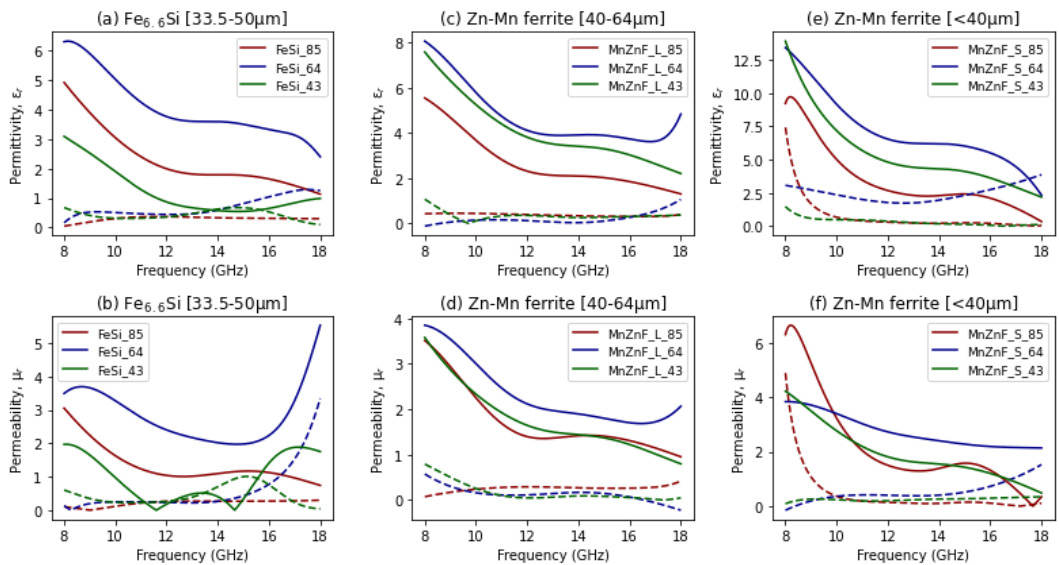

**Figure 3.** Compilation of the electromagnetic response of each sample deduced from the S-parameters after applying NRW equations. Each column shows the results for those samples

containing the same filler material. The first row shows the complex dielectric response, while the second depicts the magnetic one. Solid and dashed lines represent the real and imaginary parts of each magnitude. In brackets, the legend specified the *ffw* of each sample.

The values are decreasing with frequency in all the cases, as may be expected at this frequency range due to relaxation processes. Once the electromagnetic response of each sheet is determined, the well-known single-layer impedance model, described by equations (1) and (2), may be used to compute their theoretical absorption. Nonetheless, here we are not only relying on the theoretical prediction, but we also measured the experimental $R_L$ in an anechoic chamber under real far-field radar conditions.

$$Z = Z_0 \sqrt{\frac{\mu_r}{\varepsilon_r}} \tanh\left[\left(j\frac{2\pi f d}{c}\right)\sqrt{\mu_r \varepsilon_r}\right] \quad (1)$$

$$R_L(dB) = -20 \log \left|\frac{Z/Z_0 - 1}{Z/Z_0 + 1}\right| \quad (2)$$

As shown in Figure 4 below, the $R_L$ computed from the electromagnetic parameters shown in Figure 3 perfectly fits with the experimental data. Unfortunately, as it happens in the examples on the left-side columns of the figure, most of the samples have absorption peaks below the X-band, which makes it impossible to fit the theoretical data at this range. Nonetheless, few samples present partial peaks above 8 GHz, such as the case depicted on the right-side column in Figure 4. Despite the peaks cannot be simulated in all cases, the importance of these results must be highlighted: according to the electromagnetic parameters measured before, the computed $R_L$ makes a good prediction of the samples' absorption capacities in the X-band. Figure 4 provides just an example, taking a couple of cases, of a consistence response that we have seen for all the samples.

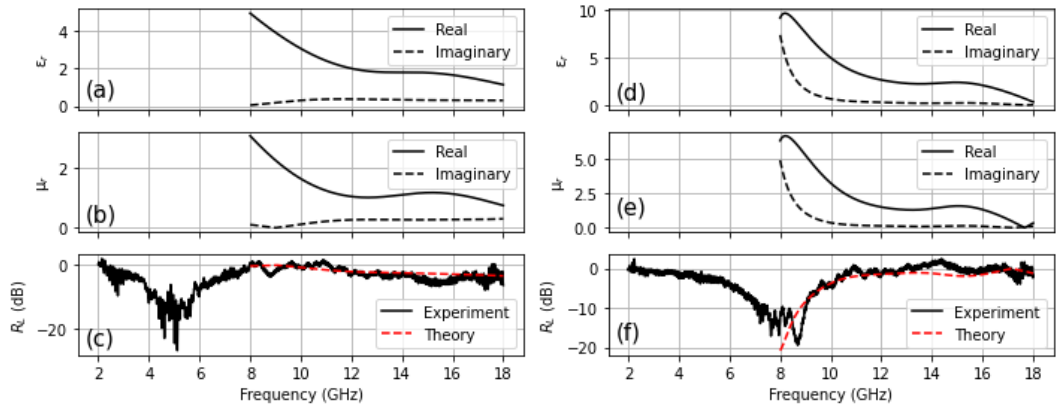

**Figure 4.** Full electromagnetic response of two samples, FeSi_85 (left-side column) and MnZnF_S_85 (right-side column). First [(a) and (d)] and second [(b) and (e)] row plots show, respectively, the complex permittivity and permeability for each sample, while the bottom one [(c) and (f)] compare the experimentally measured $R_L$ and the one computed from the electromagnetic parameters.

If we focus on the experimental absorption of the samples, Figure 5 provides a clear overview of how all the samples perform. The non-linear dependence of $R_L$ with $ff_W$ is clear for all the materials, in perfect alignment with the transmission line theory for a single layer.

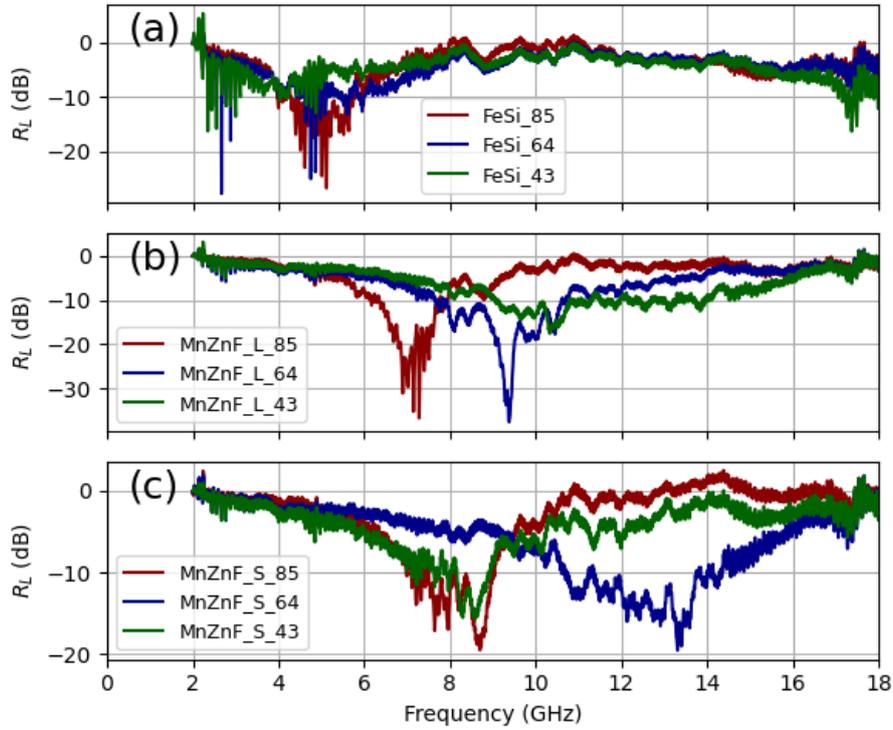

**Figure 5.** Experimental $R_L$ data was obtained in the anechoic chamber for the 9 samples. (a) shows those samples that contain $Fe_{6.6}Si$ [33.5-50μm], (b) corresponds to layers filled with Mn-Zn ferrite [40-64μm], and (c) represents samples with Mn-Zn ferrite [<40μm]. In brackets, the legend specified the $ff_W$ of each sample.

In Figure 5(a), where the $Fe_{6.6}Si$ powder is used, a broad absorption is observed in the low-frequency region of the spectrum. The $R_L$ reaches -27 dB at ~ 5 GHz for FeSi_85 and FeSi_64, while it stays at -10 dB for FeSi_43. The dependence with the $ff_W$ is evident: for the lowest $ff_W$ case (FeSi_43) there is not a meaningful absorption, while it increases when a higher load of filler is added. This change is due to the impact that a larger amount of filler has on the overall electromagnetic properties of the composite, which have a direct impact on the $R_L$. On the other hand, it must be highlighted that all three cases absorb along a wide range of frequencies: from 2 to 8 GHz.

The addition of Mn-Zn ferrite moves the absorption up to higher frequencies. The larger fraction of ferrite particles, depicted in Figure 5(b), leads to more intense peaks, while the smaller fraction shown in Figure 5(c) creates less intense but wider peaks. In the first case, we reach very meaningful attenuations of -38 dB for two of the scenarios. In the first case, despite the absorption of sample MnZnF_L_43 initially seems to be weak compared to the other two samples with higher $ff_W$ (MnZnF_L_85 and MnZnF_L_64), some remarks should be done: (i) there is a frequency shift towards higher frequencies; (ii) the peak, though it is less intense than in the other two cases, it reaches -18 dB. This is above the abovementioned -10 dB threshold typically used for defining a material as a RAM and computing its absorption bandwidth; (iii) it has the widest absorption which, considering the previous argument, makes this -18 dB very significant. The peak base covers almost all the measured range. Overall, when the $ff_W$ is increased the absorption bands become thinner but much more intense. The scenario where the most interesting combinations between peak's width and intensity are observed is when the smallest fraction of Mn-Zn ferrite is used, as it may be seen in Figure 5(c). In the case of MnZnF_S_64, for instance, there is a -20 dB peak with 3.5 GHz (10.5 to 14 GHz) bandwidth at the -10dB threshold. This peak also shows a less intense but extremely wide base, from 4 to 16 GHz. In the other two cases, absorption also reaches -20 dB with a -10 dB bandwidth of 2 GHz. For this type of filler, it is not seen a continuous change in the spectra as a function of $ff_W$. This is in perfect agreement with theory given that absorption is expected to have a nonlinear dependence on the material's permittivity and permeability, which are directly affected by $ff_W$.

The comparison made in Figure 4 suggests that the impedance model makes a good representation of the experimental data above 8 GHz; consequently, we may also use it to simulate the $R_L$ and explore a wider experimental domain. In Figure 6 we represent the computed $R_L$ between 8 and 18 GHz for layers between 0.1 and 4.0 mm.

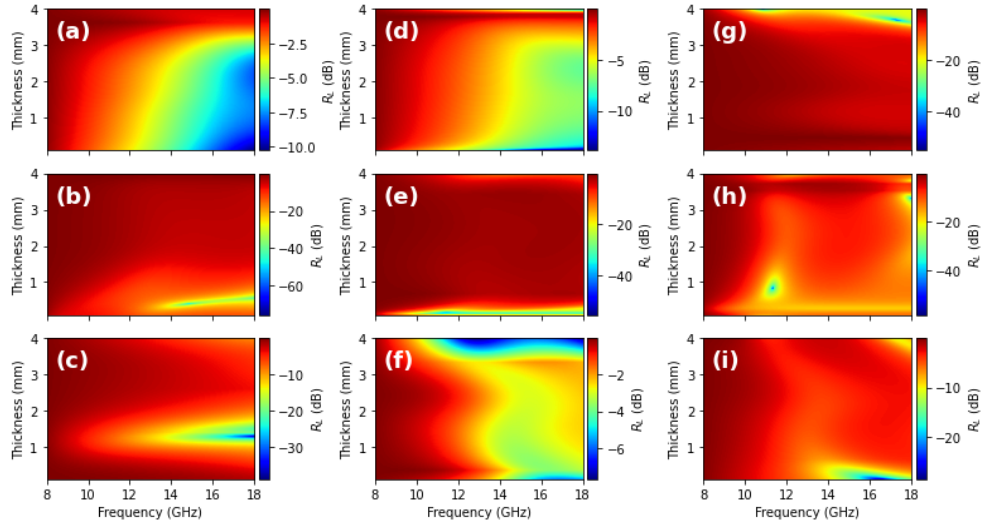

**Figure 6.** Two-dimensional color plot with the $R_L$ expected for each sample for thicknesses between 0.1 and 4.0 mm. Each column corresponds to one type of filler material, while each row corresponds to a different $ff_W$. (a) FeSi_85, (b) FeSi_64, (c) FeSi_43, (d) MnZnF_L_85, (e) MnZnF_L_64, (f) MnZnF_L_43, (g) MnZnF_S_85, (h) MnZnF_S_64, and (i) MnZnF_S_43.

As it may be observed, remarkable absorptions larger than -40 dB can be achieved with these samples between 8 and 18 GHz with a proper design. Another general aspect that must be highlighted is that the blue bands – which represent the largest absorption areas – tend to be broad in frequency for a specific thickness in most of cases. This facilitates the broadband absorption of these samples, given that the maximum peak width could be achieved by preparing layers with the optimum thickness.

To start, we analyze the impact that each type of functional filler has. FeSi (first column) is the filler that leads to the highest absorptions. Compared to MnZn ferrite, its absorption is very significant even for low $ff_W$ (bottom row of subplots), but then it is less competitive than the ferrites for the highest $ff_W$. Moving to the ferrites, we see that the smallest particle size distribution (left side column) loads to more intense absorption in all the scenarios than the largest fraction. The overall effective properties of the composite layer are affected by the unique intrinsic functional properties that each filler has, together with $ff_W$. Analyzing Figure 6 column-by-column (i.e. keeping composition constant), in all cases the highest absorption is achieved at intermediate loading conditions [$ff_W$ ~64%, subplots (b), (e), and (h)]. In these cases, however, the peaks are thinner than in the others. These results, again, support the fundamental idea that there is not a linear relationship of the electromagnetic parameters and sample thickness with $R_L$. The composite requires precise design and optimization to reach maximum broadband absorption.

Once the response of each layer is known, we now move to construct bilayer systems by combining a pair of the previous samples. The number of possible combinations between the samples listed in Table I is large and they cannot all be shown in this manuscript for evident reasons. To study these systems, we make use of the bilayer impedance model published in [22], which is described by (4) as:

$$Z = Z_0 \frac{\sqrt{\frac{\mu_1}{\varepsilon_1}}tanh\left[\left(j\frac{2\pi f d_1}{c}\right)\sqrt{\mu_1 \varepsilon_1}\right] + \sqrt{\frac{\mu_2}{\varepsilon_2}}tanh\left[\left(j\frac{2\pi f d_2}{c}\right)\sqrt{\mu_2 \varepsilon_2}\right]}{1+\sqrt{\frac{\mu_1 \varepsilon_2}{\mu_2 \varepsilon_1}}tanh\left[\left(j\frac{2\pi f d_1}{c}\right)\sqrt{\mu_1 \varepsilon_1}\right]tanh\left[\left(j\frac{2\pi f d_2}{c}\right)\sqrt{\mu_2 \varepsilon_2}\right]}$$

(3)

This model, which is directly derived from Maxwell equations, provides a physical approximation to the bilayer situation, considering each layer's electromagnetic and geometrical characteristics. Compared to the monolayer approach, here the interaction and matching of the electromagnetic constants of the two layers have a meaningful significance. It is also important to emphasize the non-linear dependence of the absorption on the electromagnetic and dimensional characteristics of the layers. The agreement between the experiments and theory is again demonstrated in Figure 7 for the combination of MnZnF_L_85 and MnZnF_L_43, as an example. In this new scenario, the main absorption peaks are always below the X-band, so they can never be simulated using the electromagnetic information of each layer. Nonetheless, the prediction is consistent with the experimental data between 8 and 18 GHz.

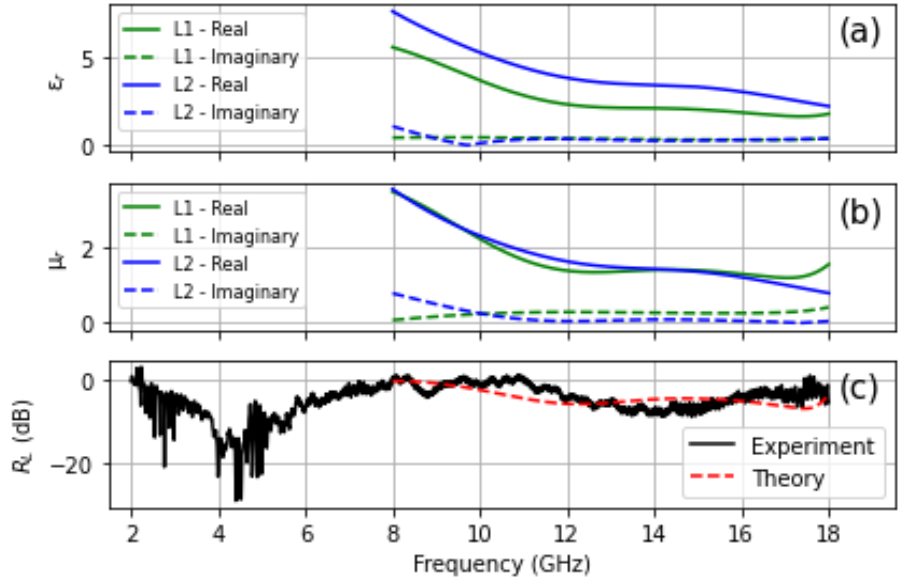

**Figure 7.** Electromagnetic properties of each of the layers [panels (a) and (b)] and comparison between the experimental $R_L$ and the calculated one using the bi-layer theory and the data above [panel (c)]. L1 represents layer 1 (MnZnF_L_85, in this case) while L2 refers to layer 2 (MnZnF_L_43).

We report some experimental results in Figure 8. Here, we show the results of the superposition of different layers systematically. We keep constant a base layer and we replace the one on top. The sample with the highest filler load of each nature (i.e., FeSi_85, MnZnF_L_85, and MnZnF_S_85) was selected as the base layer where the other was overlapped. In this figure, each row of subfigures shows the combinations keeping constant the base layer while changing the overlapped one; on the other hand, each column refers shows the obtained responses when the same sample is placed on top of different bases.

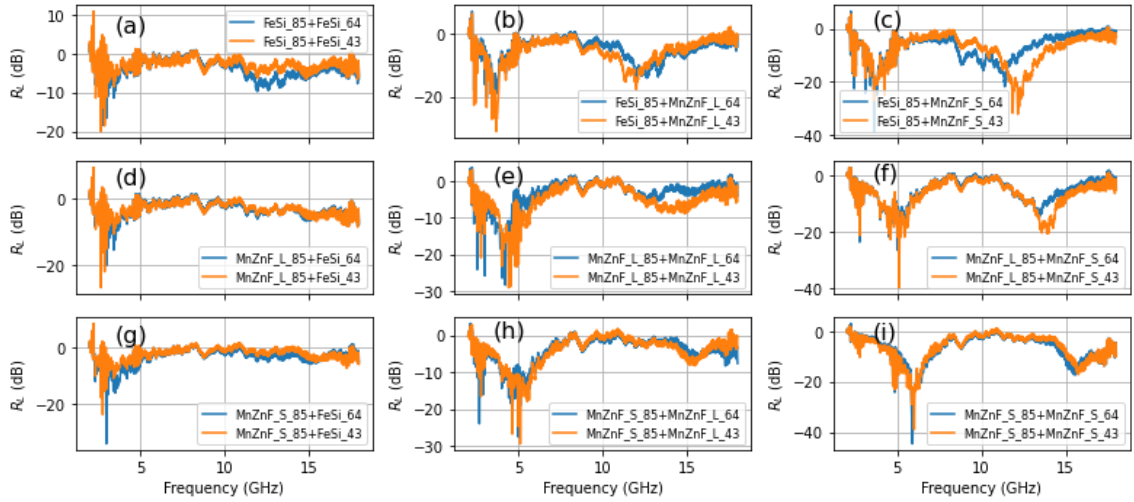

**Figure 8.** Summary of experimental $R_L$ measured on the anechoic chamber obtained by the combination of two different layers. Rows keep constant the base material, while columns maintain constant the layer added on top.

The experimental data shown in Figure 8 confirms the immense potential that bilayer systems – or multilayer, in general – have as broadband absorbers. The interaction between the two media opens a wide range of possibilities for designing novel and improved shielding systems. The differences in the results shown in Figure 5 are evident. Absorptions are increased, reaching the range of -30 or -40 dBs in multiple combinations. Nonetheless, probably the most remarkable advantage is the appearance of doublets in the spectrum. In all the 12 combinations represented in Figure 8 that contain the Mn-Zn ferrite as the additional layer (second and third columns in Figure 8) two main peaks are detected along the measured frequency range: one at ~5 GHz and a secondary one between 10 and 15 GHz. This is an interesting a powerful observation to expand the absorption range of electromagnetic absorption, which is a clear advantage of these systems for stealth technology.

Finally, we can also use the bilayer theory to simulate the response of our materials in a wider experimental range. Given that $R_L$ depends on a total of 11 variables (frequency, complex permittivity of the first layer, complex permeability of the first layer, thickness of the first layer, complex permittivity of the second layer, complex permeability of the second layer, and thickness of the second layer) the solution is not elementary even for a single combination of two layers. Taking the complex electromagnetic responses for the two layers as known, the problem reduces to three dimensions for each combination: frequency and thickness of each layer. Figure 9 shows the results for one of the investigated combinations, the result of overlapping FeSi_85 and MnZnF_S_64. Given that the solution is three-dimensional, we need to define a criterion to be able to represent the results in two dimensions and extract conclusions. In Figure 9(b) we show the maximum absolute absorption (i.e., minimum $R_L$) found in this frequency range, while Figure 9(b) represents the frequency position of the main peak in the spectrum. The selection of the optimum criteria depends on the specific requirements of the application; for broadband absorbers, we also suggest additional criteria such as the frequency bandwidth of the peak at specific $R_L$, for instance. However, it is the combination of criteria what provides a real picture of the system and its performance.

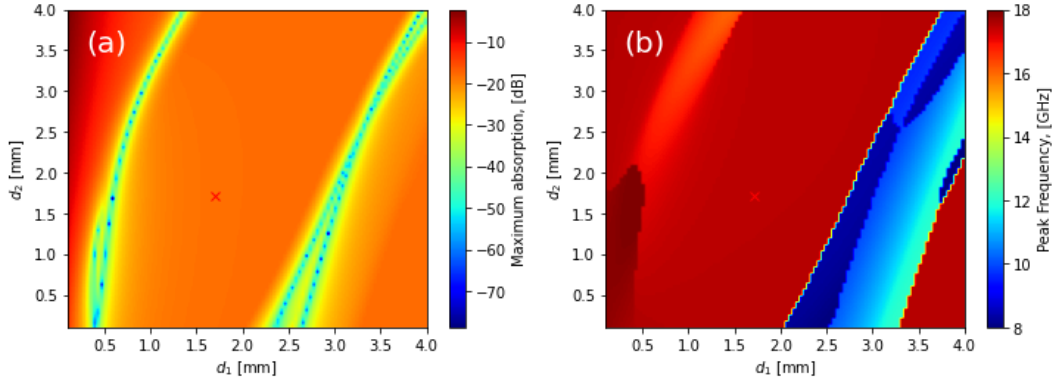

**Figure 9.** Computed $R_L$ between 8 and 18 GHz for the bilayer system made of FeSi_85 + MnZnF_S_64, computed using their electromagnetic response and the bilayer theory. (a) Represents the maximum absorption (minimum $R_L$) observed in the spectrum; and (b) shows the frequency position of the main peak. The red cross specifies where the real sample would be.

As it may be seen in Figure 9(a), the combination of FeSi_85 + MnZnF_S_64 may show a maximum absorption when one of the two layers is much thicker than the other (either being $d_1 \gg d_2$ or $d_1 \ll d_2$), represented by the blue regions. However, the blue bands are very thin, indicating that achieving optimal performance demands a high level of design precision. Nonetheless, when looking at Figure 9(b) we can see that depending on which thickness combination we choose ($d_1 \gg d_2$ or $d_1 \ll d_2$) the position of the main peak shifts. For $d_1 \gg d_2$ the maximum absorption would happen at 8 – 10 GHz, but if $d_1 \ll d_2$ the peak would move to ~ 15 GHz. These conclusions are extremely powerful for the design of novel electromagnetic shielding systems.

## 4. Conclusions

Intense and broadband radar absorption has been measured for 1.7 mm thick layers made of a matrix of PDMS filled with a high load of functional magnetic powder, either metallic or ceramic. The electromagnetic parameters (complex permittivity and permeability) have been presented for all the samples, which have a clear impact on their shielding capacities. The experimental absorption, measured in real far-field radar conditions inside an anechoic chamber, shows its strong dependence on the sample's electromagnetic properties and thickness, which must be tuned for an optimum design. The fabricated samples show reflection losses of nearly -40 dB but, according to the impedance model for single-layer systems, even larger $R_L$ could be reached with the adequate design. The experimental data also supports the fundamental concept that there is not a linear dependency of the absorption with these variables.

On the other hand, bilayer systems have also been studied in this paper. A clear modification of the individual spectra of each layer is observed due to the interaction between the two interfaces with different refractive indices. More intense and wider absorptions may be achieved by this type of combination, which also leads to the appearance of doublets under specific conditions. This has a direct impact and represents a meaningful advantage for broadband absorption applications. Our analysis concludes that the thickness combination between the two layers is a crucial parameter to achieve intense and broadband absorptions. The samples analyzed here exemplify that substantial thickness differences between the two layers may lead to high-performance radar absorbing systems.

**Acknowledgments**

This work was supported by the U.S. Air Force Office of Scientific Research (AFOSR) [grant number FA8655-22-1-7049]. Pilar Marin acknowledges the financial support from the Spanish Ministry of Economic Affairs and Digital Transformation through the project PID2021-123112OB-C21-MICIIN, to the Community of Madrid NanomagCOSt project (S2018/NMT-4321). The authors would also like to thank the enterprise PREMO S.L. for preparing the necessary materials.

**Data availability**

Data will be made available on request.

**References**

[1] K. Hussain and I. Y. Oh, "Joint Radar, Communication, and Integration of Beamforming Technology," *Electronics 2024, Vol. 13, Page 1531*, vol. 13, no. 8, p. 1531, Apr. 2024, doi: 10.3390/ELECTRONICS13081531.

[2] M. Pant and L. Malviya, "Design, developments, and applications of 5G antennas: a review," *Int J Microw Wirel Technol*, vol. 15, no. 1, pp. 156–182, Feb. 2023, doi: 10.1017/S1759078722000095.

[3] Z. Zhou, S. Liu, W. Li, T. An, and Z. Hu, "A wideband MIMO antenna with high isolation for 5G application," *International Journal of RF and Microwave Computer-Aided Engineering*, vol. 32, no. 3, p. e23004, Mar. 2022, doi: 10.1002/MMCE.23004.

[4] K. Edokossi, A. Calabia, S. Jin, and I. Molina, "GNSS-Reflectometry and Remote Sensing of Soil Moisture: A Review of Measurement Techniques, Methods, and Applications," *Remote Sensing 2020, Vol. 12, Page 614*, vol. 12, no. 4, p. 614, Feb. 2020, doi: 10.3390/RS12040614.

[5] A. Soumya, C. Krishna Mohan, and L. R. Cenkeramaddi, "Recent Advances in mmWave-Radar-Based Sensing, Its Applications, and Machine Learning Techniques: A Review," *Sensors 2023, Vol. 23, Page 8901*, vol. 23, no. 21, p. 8901, Nov. 2023, doi: 10.3390/S23218901.

[6] L. Lo Bello, G. Patti, and L. Leonardi, "A Perspective on Ethernet in Automotive Communications—Current Status and Future Trends," *Applied Sciences 2023, Vol. 13, Page 1278*, vol. 13, no. 3, p. 1278, Jan. 2023, doi: 10.3390/APP13031278.

[7] Y. Sani, R. S. Azis, I. Ismail, Y. Yaakob, and J. Mohammed, "Enhanced electromagnetic microwave absorbing performance of carbon nanostructures for RAMs: A review," *Applied Surface Science Advances*, vol. 18, p. 100455, Dec. 2023, doi: 10.1016/J.APSADV.2023.100455.

[8] S. Bao, M. Zhang, Z. Jiang, Z. Xie, and L. Zheng, "Advances in microwave absorbing materials with broad-bandwidth response," *Nano Res*, vol. 16, no. 8, pp. 11054–11083, Aug. 2023, doi: 10.1007/S12274-023-5654-6/METRICS.

[9] A. Choudhary, S. Pal, and G. Sarkhel, "Broadband millimeter-wave absorbers: a review," *Int J Microw Wirel Technol*, vol. 15, no. 2, pp. 347–363, Mar. 2023, doi: 10.1017/S1759078722000162.

[10] N. Shirke, V. Ghase, and V. Jamdar, "Recent advances in stealth coating," *Polymer Bulletin*. Springer Science and Business Media Deutschland GmbH, 2024. doi: 10.1007/s00289-024-05166-4.

[11] S. Bao, M. Zhang, Z. Jiang, Z. Xie, and L. Zheng, "Advances in microwave absorbing materials with broad-bandwidth response," *Nano Research*, vol. 16, no. 8. Tsinghua University, pp. 11054–11083, Aug. 01, 2023. doi: 10.1007/s12274-023-5654-6.

[12] L. Jin *et al.*, "Application, development, and challenges of stealth materials/structures in next-generation aviation equipment," *Applied Surface Science Advances*, vol. 19. Elsevier B.V., Feb. 01, 2024. doi: 10.1016/j.apsadv.2024.100575.

[13] U. Hwang and J. Do Nam, "Frequency-Selective Radar-Absorbing Composites Using Hybrid Core-Shell Spheres," *ACS Nano*, vol. 18, pp. 12225–12234, May 2024, doi: 10.1021/ACSNANO.4C00624/ASSET/IMAGES/LARGE/NN4C00624_0005.JPEG.

[14] S. Wei *et al.*, "Constructing and optimizing epoxy resin-based carbon Nanotube/Barium ferrite microwave absorbing coating system," *Mater Res Bull*, vol. 179, p. 112928, Nov. 2024, doi: 10.1016/J.MATERRESBULL.2024.112928.

[15] A. Castellano-Soria, E. Navarro, J. López-Sánchez, and P. Marín, "A novel methodology for designing Mono/Bi-slab X-band microwave absorbers of Carbon-Powder composites," *Mater Des*, vol. 238, p. 112641, Feb. 2024, doi: 10.1016/J.MATDES.2024.112641.

[16] A. G. Gorriti, P. Marin, D. Cortina, and A. Hernando, "Microwave attenuation with composite of copper microwires," *J Magn Magn Mater*, vol. 322, no. 9–12, pp. 1505–1510, May 2010, doi: 10.1016/J.JMMM.2009.07.085.

[17] P. Marín, D. Cortina, and A. Hernando, "Electromagnetic wave absorbing material based on magnetic microwires," *IEEE Trans Magn*, vol. 44, no. 11 PART 2, pp. 3934–3937, 2008, doi: 10.1109/TMAG.2008.2002472.

[18] P. G. B. Gueye, J. L. Sánchez, E. Navarro, A. Serrano, and P. Marín, "Control of the Length of Fe73.5Si13.5Nb3Cu1B9 Microwires to Be Used for Magnetic and Microwave Absorbing Purposes," *ACS Appl Mater Interfaces*, vol. 12, no. 13, pp. 15644–15656, 2020, doi: 10.1021/acsami.9b21865.

[19] J. Calvo-de la Rosa, J. Tejada, and A. Lousa, "Structural and impedance spectroscopy characterization of Soft Magnetic Materials," *J Magn Magn Mater*, vol. 475, pp. 570–578, 2019, doi: 10.1016/j.jmmm.2018.11.085.

[20] J. Calvo-De La Rosa, J. Vanacken, V. V. Moshchalkov, and J. Tejada, "Pulsed Magnetic Field Experiments in SMM and SMC Materials: New Questions about Its Properties and Applications," *IEEE Trans Magn*, vol. 57, no. 6, pp. 18–21, 2021, doi: 10.1109/TMAG.2021.3068059.

[21] J. Calvo-De La Rosa, J. Manel Hernàndez, A. García-Santiago, J. Maria Lopez-Villegas, and J. Tejada, "Barium Hexaferrite-based nanocomposites as random field magnets for microwave absorption." doi: https://doi.org/10.48550/arXiv.2402.14324.

[22] J. Calvo-de la Rosa *et al.*, "New Approach to Designing Functional Materials for Stealth Technology: Radar Experiment with Bilayer Absorbers and Optimization of the Reflection Loss," *Adv Funct Mater*, p. 2308819, Oct. 2023, doi: 10.1002/adfm.202308819.

[23]  A. M. Nicolson and G. F. Ross, "Measurement of the Intrinsic Properties Of Materials by Time-Domain Techniques," *IEEE Trans Instrum Meas*, vol. 19, no. 4, pp. 377–382, 1970, doi: 10.1109/TIM.1970.4313932.